\documentclass[showpacs,preprintnumbers,amsmath,amssymb,superscriptaddress,prb,preprint]{revtex4}
\usepackage{epsfig}
\usepackage{graphicx}

\newcommand{\beq}[1]{
\begin{equation}
\label{e#1} }

\newcommand{\eeq}{
\end{equation}
}

\begin{document}

\title{Enhanced annealing, high Curie temperature and low-voltage gating in (Ga,Mn)As: A surface oxide control study}
\author{K. Olejn\'{\i}k}
\affiliation{Institute of Physics ASCR v.v.i., Cukrovarnick\'a 10, 162 53
Praha 6, Czech Republic}

\author{M.~H.~S.~Owen}
\affiliation{Hitachi Cambridge Laboratory, Cambridge CB3 0HE, United Kingdom}
\affiliation{Microelectronics Research Centre, Cavendish Laboratory,
University of Cambridge, CB3 0HE, United Kingdom}

\author{V. Nov\'ak}
\affiliation{Institute of Physics ASCR v.v.i., Cukrovarnick\'a 10, 162 53
Praha 6, Czech Republic}

\author{J. Ma\v{s}ek}
\affiliation{Institute of Physics ASCR v.v.i., Na Slovance 2, 182 21 Praha
8, Czech Republic}

\author{A.~C.~Irvine}
\affiliation{Microelectronics Research Centre, Cavendish Laboratory,
University of Cambridge, CB3 0HE, United Kingdom}

\author{J.~Wunderlich}
\affiliation{Hitachi Cambridge Laboratory, Cambridge CB3 0HE, United Kingdom}

\author{T.~Jungwirth}
\affiliation{Institute of Physics ASCR v.v.i., Cukrovarnick\'a 10, 162 53
Praha 6, Czech Republic} \affiliation{School of Physics and
Astronomy, University of Nottingham, Nottingham NG7 2RD, United Kingdom}
\date{\today}
%

%

\maketitle

{\bf
(Ga,Mn)As and related diluted magnetic semiconductors\cite{Matsukura:2002_a,Jungwirth:2006_a} play a major role in spintronics research because of their potential to combine ferromagnetism and semiconducting properties in one physical system. Ferromagnetism requires $\sim 1-10$\% of substitutional Mn$_{\rm Ga}$. Unintentional defects  formed during growth at these high dopings significantly suppress the Curie temperature. We present experiments in which by etching the (Ga,Mn)As surface oxide we achieve a dramatic reduction of annealing times necessary to optimize the ferromagnetic film after growth, and report Curie temperature of 180~K at approximately 8\% of Mn$_{\rm Ga}$. Our study elucidates the mechanism controlling the removal of the most detrimental, interstitial Mn defect. The limits and utility of electrical gating of the highly-doped (Ga,Mn)As semiconductor are not yet established; so far electric-field effects have been demonstrated on magnetization with tens of Volts applied on a top-gate, field effect transistor structure.\cite{Chiba:2006_b} In the second part of the paper we present a back-gate, n-GaAs/AlAs/GaMnAs transistor operating at a few Volts. Inspired by the etching study of (Ga,Mn)As films we apply the oxide-etching/re-oxidation procedure to reduce the thickness (arial density of carriers) of the (Ga,Mn)As  and observe a large enhancement of the gating efficiency. We report gatable spintronic characteristics on a series of anisotropic magnetoresistance measurements.

}

(Ga,Mn)As samples employed in our study were grown by low-temperature ($\sim 200^{\circ}$C) molecular beam epitaxy. While the non-equilibrium growth is necessary for avoiding Mn precipitation at concentrations well above the solubility limit, it inevitably leads to the formation of a large number of metastable defects.\cite{Matsukura:2002_a,Jungwirth:2006_a} It is now firmly established that the most important unintentional impurities are interstitial Mn  atoms\cite{Maca:2002_a,Yu:2002_a} which act as compensating donors and reduce the magnetic moment by forming antiferromagnetically coupled pairs with substitutional Mn$_{\rm Ga}$ acceptors. In highly doped (Ga,Mn)As the fraction of interstitial Mn becomes significant\cite{Jungwirth:2005_b} with a largely detrimental effect on ferromagnetism and electrical conduction.

Interstitial Mn impurities can be removed by post-growth annealing at temperatures close to the growth temperature, while keeping the substitutional Mn$_{\rm Ga}$ ions in place.\cite{Yu:2002_a,Edmonds:2002_b,Chiba:2003_b,Ku:2003_a,Stone:2003_a}
A number of complementary experiments have concluded that interstitial Mn ions out-diffuse during annealing towards the (Ga,Mn)As surface. Nevertheless, whether the mechanism limiting the diffusion has its origin in the (Ga,Mn)As bulk or in the surface oxide has remained a controversial open problem.\cite{Stone:2003_a,Chiba:2003_b,Kirby:2004_a,Malfait:2004_a,Edmonds:2004_a,Adell:2004_b,Sadowski:2006_a,Kirby:2006_a,Holy:2006_a}
Figures~1 and 2 provide a resolution to this question and show that it has important consequences for the efficiency of the post-growth optimization  of magnetic properties of the material.

Figure~1(a) shows SQUID magnetization measurements on a 35~nm (Ga,Mn)As film with 11\% nominal Mn-doping. The annealing was performed in air by placing the sample on a hot plate with temperature 200$^\circ$C. Two different annealing procedures are compared for samples cleaved from the same wafer which was rotated during growth to ensure homogeneity across the epilayer. Beside the standard continuous annealing, a sequence of discrete etch-anneal steps has
been applied, each consisting of 30~s etching in 30\% HCl, rinsing in water, and subsequent annealing at 200$^{\circ}$C for 5~min. With continuous annealing, the as-grown Curie temperature $T_c=$85~K is increased to $T_c=$137~K after 15~min of annealing  and a maximum $T_c=176$~K is reached in 100~min. By employing the etch-anneal technique, $T_c=170~K$ is reached in 15~min, and a saturated value of $T_c=180$~K is obtained in 40~min. We see a large reduction of the annealing time due to surface-oxide etching. The procedure also yields a sizable increase of the maximum attainable Curie temperature in the material. Our $T_c=180$~K represents an increase of the record Curie temperature achieved in (Ga,Mn)As to date\cite{Jungwirth:2005_b} by 7~K and is the highest $T_c$ reported for uniform, carrier-mediated (III,Mn)V ferromagnet. The Mn$_{\rm Ga}$ density for the sample is estimated to 8\% and the $T_c$ value, when compared with  a series of our other as-grown and annealed materials of a wide range of Mn dopings, shows no indication of reaching a Curie temperature limit in (Ga,Mn)As.

In Figure~1(b) we present a detailed investigation of the etch-anneal process. To demonstrate that our findings are generic we show results for a standard, 7\% Mn-doped 50~nm thick (Ga,Mn)As epilayer. In the main panel of Figure~1(b) we compare two sets of SQUID Curie temperature measurements as a function of annealing time, one for continuous annealing and the other for the etch-anneal procedure. In the inset, we compare these two measurements with annealing of the sample exposed to a single etching before the first annealing step.  The presence of the surface oxide layer clearly plays the crucial limiting role in the (Ga,Mn)As annealing process. By repeating several etch-anneal steps instead of annealing continuously, the post-growth treatment time for reaching maximum $T_c$, around  150~K in this particular wafer, is slashed by a factor of 10. Furthermore, the inset indicates that the oxide layer formed after growth, which very rich in Mn as discussed below, and the oxide layer regrown on a clean (Ga,Mn)As surface during annealing are comparably efficient in blocking the interstitial Mn removal.

Without passivation of out-diffused Mn-ions by oxidation (or nitridation, etc.), the diffusion of the interstitial Mn towards the (Ga,Mn)As free surface is inhibited by  the formation of an electrostatic barrier.\cite{Stone:2003_a,Chiba:2003_b,Kirby:2004_a,Malfait:2004_a,Adell:2004_b,Sadowski:2006_a} From this perspective the key role of the surface oxide in the annealing process is plausible as it might prevent the Mn-ions and free oxygen from interacting with each other. It is a generic feature, related to the shorter (by $\sim$20\%) metal-oxygen bond length compared to GaAs, less interstitial space, and amorphous structure, that the impurity diffusion channels through the oxide are effectively blocked.\cite{Ahman:1996_a} To investigate the structure of the oxide layer and the Mn-ion passivation process in more detail we performed a series of complementary x-ray photoemission spectroscopy (XPS), magnetization, and transport measurements.

The angular dependent XPS study was done on a 50~nm (Ga,Mn)As epilayer with  8\% Mn-doping from which we prepared four different samples: a piece of an as-grown material, a sample once etched after growth and unannealed, a piece etched and annealed for 5~min, and another sample etched and annealed for 20~min. Ga-$3d$, Mn-$2p$, and O-$1s$ photoelectron line intensities were recorded with the angle-dependent probing depth upto 3~nm, and we used SESSA software package to simulate the XPS data. The fit of the measured O-$1s$/Ga-$3d$ ratios yields  9~\AA\, thickness of the oxidized (Ga,Mn)As layer in the as-grown material; in the etched unannealed sample, measured 30~min after etching, the fitted oxidized layer thickness is 5.3~\AA (comparable to results of XPS measurements in pure GaAs.\cite{Storm:1994_a}). The angular dependencies of the Mn-$2p$/Ga-$3d$ ratio for the four studied samples are plotted in Figure~2(a). The best fit for the as-grown material is obtained assuming 8.8\% Mn concentration in the unoxidized (Ga,Mn)As, which is consistent with the nominal Mn-doping, and a 44\% Mn concentration in the 9~\AA\, surface oxide. In the etched unannealed samples the same fitted values of 8.8\% Mn are found in the 5.3~\AA\, oxidized layer and in the (Ga,Mn)As film underneath. This indicates that after etching off the Mn-rich surface layer formed during growth, a thinner layer of (Ga,Mn)As is oxidized without any marked redistribution of Mn throughout the (Ga,Mn)As epilayer.

The analysis of the etched and 20~min annealed sample, which shows a sizable reemergence of Mn at the surface due to annealing, is detailed in Figure~2(b). To fit the corresponding XPS data we considered a Mn rich layer on top of the oxide, a uniform distribution of Mn in the Mn-rich oxide, and a Mn-rich layer at the interface of the oxide and unoxidized (Ga,Mn)As (see Figure~2(c)). Clearly the last model provides the best fit to the measured data with an average thickness of the Mn-rich interfacial layer of 2~\AA. We conclude that the Mn passivation takes place at the bottom of the oxide layer by oxygen which has diffused through the oxide and reached the interface with (Ga,Mn)As. Note that the picture is consistent with previous studies of GaAs oxidation which found a substantially smaller diffusivity through the oxide film of metal atoms as compared to oxygen.\cite{Wolan:1998_a}


Complementary estimates of the oxide thickness based on measured reduction of the magnetic moment and sheet conductance, which can be readily performed in highly-doped thin (Ga,Mn)As films, are consistent with the XPS analysis. In Figure~2(d) we show two subsequent measurements of the time-dependent room-temperature  resistance of a nominally  10~nm, 8\% Mn-doped (Ga,Mn)As wafer. Each experiment started with etching the surface oxide and then, after two minutes, we began monitoring for several hours the time evolution of the resistance due to re-oxidation of the (Ga,Mn)As surface. Oxide layer thicknesses were obtained by assuming ohmic conduction in the unoxidized (Ga,Mn)As.  In a few minutes after etching, a several Angstrom thick oxide layer quickly forms at the surface, followed by a significantly lower-rate oxidation process in which the thickness of the oxide reaches 1~nm in several hours.  Oxide etching and subsequent re-oxidation can therefore be used for a controllable thinning of (Ga,Mn)As films in sub-nm steps when studying numerous scaling characteristics of ferromagnetic semiconductors with the film thickness and arial density of carriers. Gating of ultrathin (Ga,Mn)As epilayers, discussed in the remaining paragraphs, is one example where this simple table-top technique can be utilized.

In Figures~3(a) and (b) we show schematics of our (Ga,Mn)As field effect transistor (FET) heterostructure together with the top-view of the Corbino geometry of the source and drain electrodes used for transport measurements. A heavily doped epitaxial n-GaAs buffer layer grown on an n-type substrate forms the gate electrode which is separated from the (Ga,Mn)As film by a 20~nm thick AlAs barrier. An appreciable gate action is expected for hole depletion/accummulation of $\sim 10^{20}$~cm$^{-3}$ which in our geometry is achieved at gate voltages of a few Volts for (Ga,Mn)As thicknesses of 5~nm and lower. This is about an order of magnitude smaller voltage as compared to previous experiments in top-gate (Ga,Mn)As/Al$_2$O$_3$/metal FETs.\cite{Chiba:2006_b} Furthermore, because of the free (Ga,Mn)As surface in our back-gate FET, the structure is directly compatible with the oxide-etching technique.

Figures~3(b) and (c) show two gating characteristics of the resistance of one physical device in which the Corbino contacts were first patterned by optical lithography on a 4~nm thick (plus a 1~nm oxidized surface layer), 8\% Mn-doped (Ga,Mn)As. After measuring this device, the (Ga,Mn)As layer was further etched to approximately 3~nm without removing the contacts, and again measured. In both cases we find a decrease/increase of the (Ga,Mn)As channel conductance upon depletion/accummulation induced by gating. (Note that the asymmetry between positive and negative gate voltage characteristics reflects the p-n junction nature of our FET.) The gating efficiency is enhanced in the thinned sample, consistent with the reduced arial density of carriers. We attribute the large amplification of this effect with decreasing temperature to the vicinity of the metal-insulator transition in the studied (Ga,Mn)As film.

Detailed measurements of gate-dependent magnetoresistance (MR) characteristics were performed in the 4~nm (Ga,Mn)As transistor to investigate its utility as a three-terminal spintronic device and as a new tool for exploring the dependence of magnetotransport in a ferromagnet on carrier concentration. The source-gate leakage current in the device is negligible compared to the source-drain current for gate voltages in the range from -2.5 to +3~V at temperatures larger than 4~K. Examples of the rich MR phenomenology we observe are shown in Figure~4. In panel (a) we plot MR curves for magnetic fields oriented nearly perpendicular to the (Ga,Mn)As plane. The high-field gate dependent MRs are discussed below. In this plot we highlight (see the inset) the low-field switching event in the in-plane magnetization component, reflected as a peak on the MR traces at about 120~mT whose position moves by 10's of mT between depletion and accumulation. In-plane magnetization rotation experiments shown in Figure~4(b) illustrate that besides the  magnetic anisotropies  also the in-plane anisotropic magnetoresistance (AMR) is modified by the gate. In the Corbino geometry with in-plane field the non-crystalline AMR component given by the relative angle between current and magnetization is expected to average to zero so the measured signal corresponds to the typically more subtle crystalline AMR terms.\cite{Rushforth:2007_a} We find that even these terms are sensitive to depletion or accumulation of holes achieved in our (Ga,Mn)As FET structure. Larger crystalline AMR we observe for depletion is qualitatively consistent with theoretically expected trends for magnetocrystalline anisotropy coefficients in (Ga,Mn)As.\cite{Rushforth:2007_a}

Stronger gate-dependence of MR signals is observed when magnetization is rotated from the in-plane, magnetic easy direction towards the out-of-plane hard axis. In Figure~4(c) we compare field-sweep measurements for in-plane magnetic fields and for the perpendicular-to-plane fields. Since magnetization is expected to remain parallel with the field in the former experiment, the trace is interpreted as the isotropic negative MR for which we find only a very week dependence on the gate voltage. By subtracting the isotropic MR part from the curves measured in the perpendicular-to-plane magnetic field sweeps we obtain the out-of-plane AMRs (see Figure~4(d)), i.e., the difference between resistances for the in-plane magnetization orientation at low fields and for the out-of-plane orientation at high saturating fields. Unlike the in-plane AMR, here the non-crystalline AMR component is present even in the Corbino geometry and we see that the AMR decreases from 5\% at -2.5~V (accumulation) to 4\% at 3~V (depletion), again qualitatively consistent with theoretically expected trends for the non-crystalline AMR.\cite{Rushforth:2007_a} Note that part of the out-of-plane MR may be due to the anomalous Hall effect whose contribution cannot be directly separated from the AMR in the Corbino device. Nevertheless, since the anomalous Hall (and normal Hall at 1~T) resistance is typically much smaller than the sheet resistance in (Ga,Mn)As, the AMR is likely the dominant contribution to MR traces in Figure ~4(d).

To summarize we have presented several advancements  towards functional (Ga,Mn)As ferromagnetic semiconductors for spintronics. By introducing the etch-anneal post-growth treatment we have significantly enhanced the out-diffusion of compensating interstitial Mn defects and elucidated the basic mechanism controlling their passivation at the surface. The technique helped us to prepare (Ga,Mn)As material with the highest Curie temperature reported to date within the family of bulk (III,Mn)V carrier-mediated ferromagnets.  Electrical gating of ultra-thin (Ga,Mn)As films was  explored on a back-gate, all-semiconductor (Ga,Mn)As FET. The device operates at gate voltages of a few Volts. Its free (Ga,Mn)As surface geometry allows for a controllable  thinning of the (Ga,Mn)As layer in sub-nanometer steps by etching and re-oxidation, resulting in a significant enhancement of the gating efficiency. Our observations of  gate-dependent crystalline and non-crystalline AMR responses open new ways for exploring MR effects in ferromagnets and prototype realizations of three-terminal spintronic devices.

We acknowledge fruitful collaborations with R.~P.~Campion, M.~Cukr, B.~L.~Gallagher, M.~Mary\v{s}ko, and J.~Zemek, and support from EU Grant IST-015728, from Grant Agency and Academy of Sciences and Ministry of Education of the Czech Republic Grants FON/06/E001, FON/06/E002, AV0Z1010052, KAN400100652, and LC510.


\begin{figure}[h]

\vspace*{-1.5cm}
\epsfig{width=0.78\columnwidth,angle=90,file=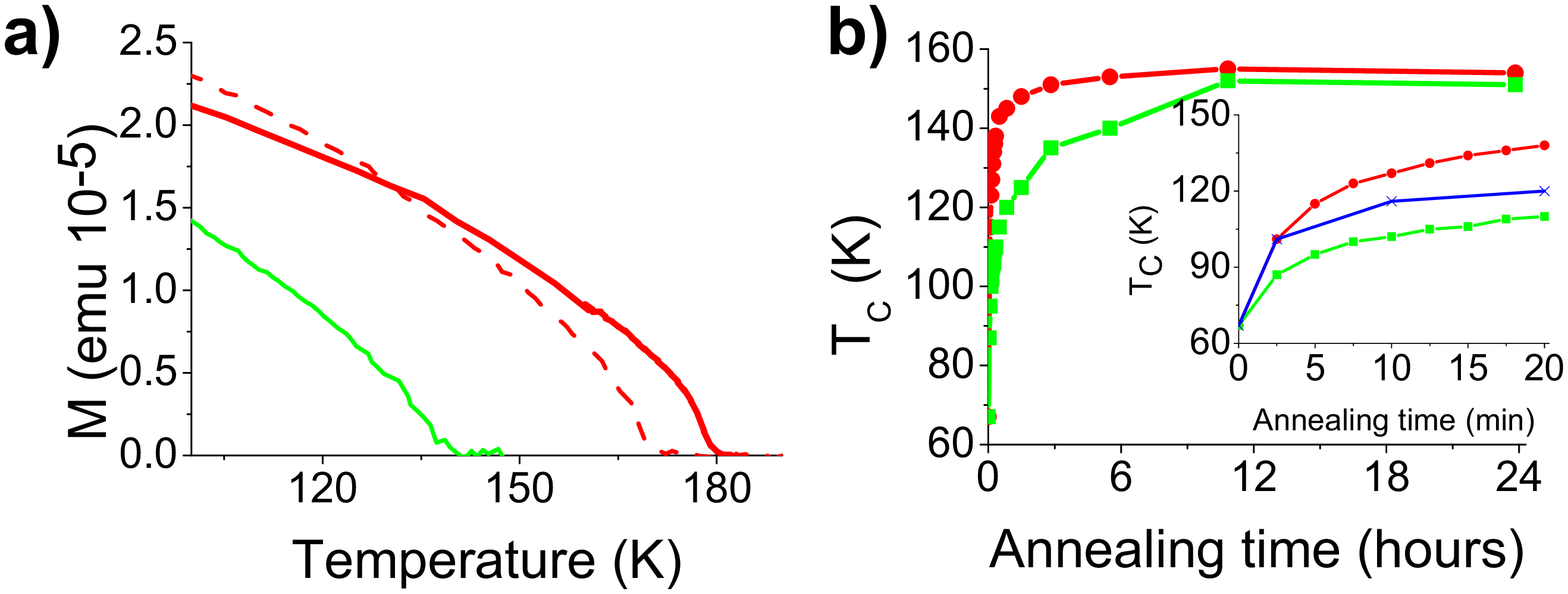}

\vspace*{-1.5cm}
\caption{(a) Remanent magnetization measured by a superconducting quantum interference device (SQUID) as a function of temperature for a 15~min continuously annealed sample (green curve) and for samples treated with the etch-anneal procedure for 15~min (red dashed line) and 40~min (red solid line). All samples were prepared from a 35~nm, 11\%-doped (Ga,Mn)As wafer.(b) Curie temperature as a function of the post-growth treatment time for the continuous annealing (green symbols), for the etch-anneal procedure including etching after each annealing interval (red symbols), and for continuous annealing following a single etching of the as-grown film (blue symbols in the inset). A standard, 50~nm 7\%-doped (Ga,Mn)As wafer was used for this study.}
\label{f1}
\end{figure}

\begin{figure}[h]
\epsfig{width=0.7\columnwidth,angle=90,file=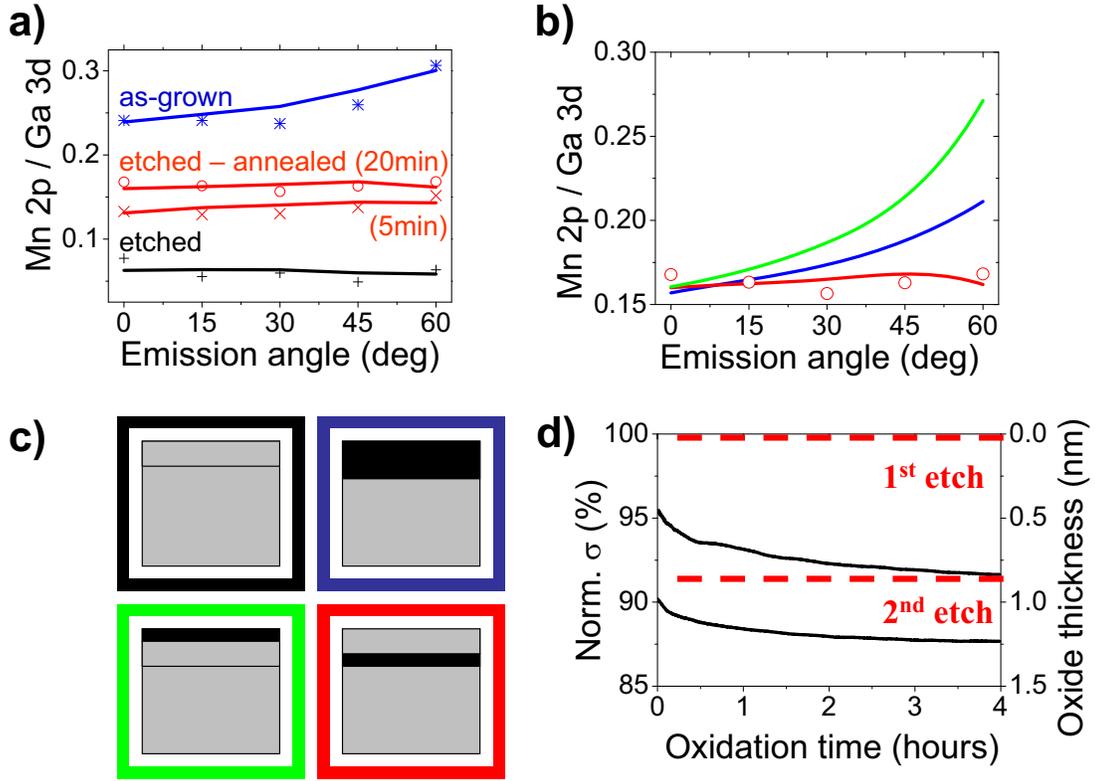}
\caption{(a) Angle-resolved XPS measurements of the ratio of Mn-$2p$ and Ga-$3d$ photoelectron line intensities on an as-grown, etched, and etched-annealed 50~nm, 8\% Mn-doped (Ga,Mn)As. Color coding of the symbols and fitted lines correspond to the models sketched in panel (c) of Mn-distribution in the oxidized and unoxidized part of the (Ga,Mn)As film assumed in the fitting. Black color corresponds to Mn uniformly distributed throughout the (Ga,Mn)As and the oxide; blue color to Mn-rich region covering uniformly the oxide layer; green color to Mn-rich surface layer; red color to Mn-rich interfacial layer between the  oxidized and unoxidized parts of (Ga,Mn)As. (b) XPS measurements on the etched and 20~min annealed sample (red symbols) and fitted curves with the color coding corresponding to Mn-distribution models of panel (c). (d) Two subsequent measurements, starting from an as-grown 10~nm, 8\% Mn-doped (Ga,Mn)As wafer, of time dependent normalized conductance recorded from 2~min after etching. Oxide thicknesses are recalculated assuming ohmic conduction of the unoxidized (Ga,Mn)As layer.}
\label{f2}
\end{figure}

\begin{figure}[h]
\epsfig{width=0.7\columnwidth,angle=90,file=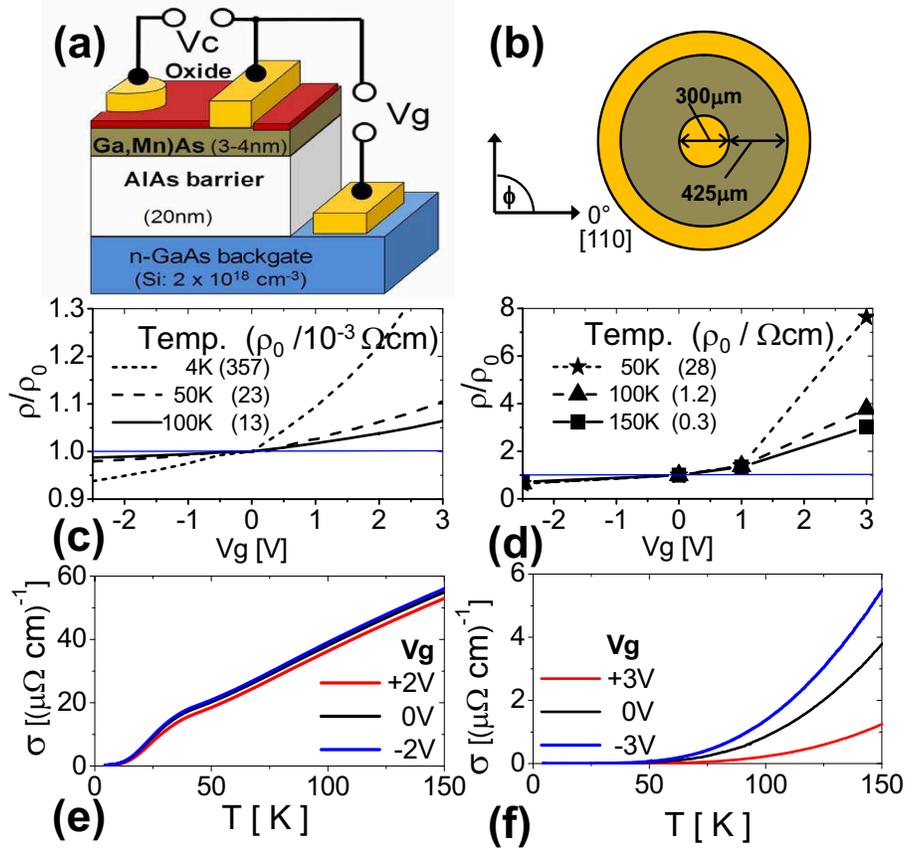}
\caption{(a) Schematic side-view of the (Ga,Mn)As FET structure. (b) Schematic top view of the source and drain electrodes arranged in a Corbino disk geometry. (c) Gate voltage dependence of the source-drain resistivity relative to the resistivity at zero gate voltage for the FET device with  4~nm thick (Ga,Mn)As (plus a 1~nm oxidized surface layer) measured at different temperatures. Same as (c) after thinning the (Ga,Mn)As layer to 3~nm by etching. (e) Temperature dependent relative resistivity in the 4~nm (Ga,Mn)As device at different gate-voltages. (f) Same as (e) for the 3~nm (Ga,Mn)As device.}
\label{f3}
\end{figure}

\begin{figure}[h]
\epsfig{width=0.7\columnwidth,angle=90,file=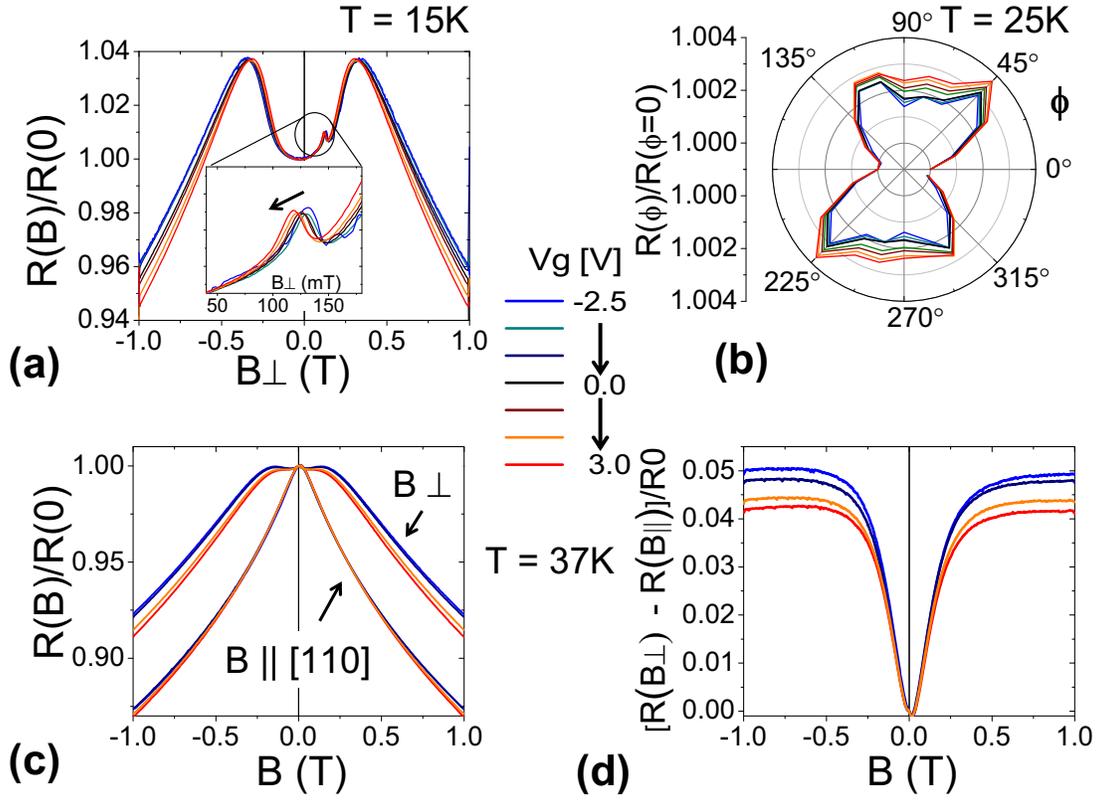}
\caption{(a) Source-drain MR for field aligned nearly perpendicular to the (Ga,Mn)As plane measured in the 4~nm (Ga,Mn)As FET device. (b) AMR measured in a rotating in-plane saturation field. Angle $\phi$ is measured from the [100] crystallographic axis. (c) Comparison of in-plane and out-of-plane MR measurements. (d) Difference between in-plane and out-of-plane magnetoresistance curves corresponding to the out-of-plane AMR. The measurement temperatures are indicated in each panel.}
\label{f4}
\end{figure}
\end{document}